\documentclass[[preprint]{aastex701}

\newcommand\ergcms{erg\,cm$^{-2}$\,s$^{-1}$}
\graphicspath{{./}{figures/}}

\shorttitle{The secular evolution of planetary nebula IC~418}
\shortauthors{A. Zijlstra \&\ Parker}

\newcommand{\ot}{[O\,{\sc iii}]}

\begin{document}

\title{The secular evolution of planetary nebula IC~418 and its implications for carbon star formation}

\author[0000-0002-3171-5469]{Albert A. Zijlstra}
\affiliation{Jodrell Bank Centre for Astrophysics,  Department of Physics and Astronomy\\ 
The University of Manchester, Oxford 
Road, M13 9PL, Manchester, UK}
\email{albert.zijlstra@manchester.ac.uk}

\correspondingauthor{Albert A. Zijlstra}

\author[0000-0002-2062-0173]{Quentin A. Parker}
\affiliation{The Laboratory for Space Research, Faculty of Science, \\ The
  University of Hong Kong, Cyberport 4, Hong Kong}
  \email{quentinp@hku.hk}



\begin{abstract}
\noindent
The rate of stellar evolution can rarely be measured in real time. The fastest evolution (excluding event-driven evolution), 
where stars may evolve measurably over decades, is during the post-AGB phase. In this paper we provide direct evidence for such a case. A secular, linear, factor of $\sim$2.5 increase is found in the strength of the \ot\ lines relative to H$\beta$ over an 130 year period in the young, well-known, low excitation planetary nebula IC\,418.  The increase is caused by the rising temperature of the central star. We use photo-ionization models to derive a model dependent heating rate for the central star in the range $15$--$42$\,K\,yr$^{-1}$. These derived heating rates are very sensitive to the stellar mass, and
yield a central-star mass of $\sim 0.560$--$0.583\,\rm M_\odot$. Initial-final mass relations based on the Miller-Bertolami models give a progenitor main-sequence mass of 1.25--1.55\,M$_\odot$. IC\,418 is a carbon rich planetary nebula and its central star, 
HD\,35914, has evolved from an AGB carbon star. This result shows that carbon star formation at solar metallicity 
extends to these low masses. This is lower than commonly assumed and suggests that post-AGB evolution 
may be slower than recent post-AGB models predict.
\end{abstract}




\section{Introduction}
\label{sec:intro}
Planetary nebulae (PNe) are a short-lived phase in the late evolution of low to intermediate-mass stars 
\citep{2022PASP..134b2001K}, after the star has terminated the Asymptotic Giant Branch (AGB) phase. 
On the AGB,  the progenitor 
star loses 40\%\  to 80\%\ of its mass \citep{Hofner2018} over several hundred thousand years via pulsations and dust-driven 
winds. The remnant proto-white-dwarf core heats up to temperatures $T \sim 10^5$\,K before nuclear burning ceases. Thereafter, 
the stellar luminosity rapidly drops as the stellar core contracts and enters the white dwarf (WD) cooling track.  The 
expanding nebula from the ejected envelope becomes ionized once the residual central star (CSPN) reaches $T = 20$--25\,kK.  
The onset of Balmer line emission signals the brief PN phase which typically lasts for 5,000 to 25,000 years \citep[e.g.,][]
{Ali2015}. 

Ionisation increases with CSPN temperature evolution during the PN phase. Between an CSPN $T_{\rm eff}$ of 30 and 45\,kK the 
integrated \ot\  and [Ne\,{\sc iii}] emission increase by more than an order of magnitude, while the integrated, higher 
excitation He\,{\sc ii} emission strongly increases between 60 and 80kK. The rate of CSPN temperature increase is a very 
strong function of stellar mass \citep{Bloecker1995b}. This can give significant evolution on decadal timescales detectable 
photometrically and spectroscopically. If accurate spectrophotometry over several decades exists, such 
changes can be observed and quantified \citep{Hajduk2015}. This is a powerful technique, since if the CSPN heating rate can 
be measured, the CSPN mass can be inferred to much higher accuracy than is possible from its luminosity \citep{Gesicki2007, 
Bloecker1995b}. With the current mass and using the initial-final mass relation, e.g. \citet{2018ApJ...866...21C}, the 
progenitor star's mass and age can be derived, yielding key information on stellar evolution.

The CSPN fades optically because most of its energy is increasingly emitted at shorter wavelengths as the star heats up on it's way to becoming a WD. Secular changes in the star's visual magnitude are expected to be small on decadal timescales, 
whereas spectroscopic changes in the nebula can be more pronounced and easier to detect.

PNe spectroscopy is 150+ years old \citep{Huggins1864}, so uncovering evolution is possible if sufficient observations for 
suitable PNe exist. However, very few PNe have decent spectral coverage, repeated over time periods sufficient to reveal 
secular evolution of their CSPN. The young, high surface brightness PN\,IC\,418 is one such rare example with a long, 130 
year record of spectroscopic observation as one of the first PN identified. Furthermore, it is in a low ionisation state 
where CSPN evolution over relatively short timescales is easier to discern. This makes IC\,418 uniquely suitable for studying 
real-time spectroscopic evolution of a PN. 

Here, we present strong observational evidence for a significant factor $\sim2.5$ increase in the observed 
\ot$\lambda$5007/H$\beta$ ratio in this bright PN as revealed by carefully vetting, and in some cases re-measuring, diverse, archival, spectroscopic data available over 130 years, and including our own more recent observations.\footnote{In this paper, "\ot/H$\beta$" refers specifically to the \ot\,$\lambda$5007\AA/H$\beta$ ratio.}

\subsection{PNe with evidence for evolution}

The few PNe that have coverage over sufficient time to reveal secular evolution include NGC\,7027 \citep[][from radio flux measurements]{zijlstra2008}, Hen\,2-260 \citep{2014A&A...567A..15H}, 
NGC\,6572 \citep{Arkhipova2014} and IC\,4997 \citep{ Aller1966,
  Feibelman1992, Kostyakova2009} (all from spectroscopy). NGC\,6572 in particular has evidence of an increase in the observed \ot/H$\beta$ ratio between 1938 and 2013 by as much as $\sim$30\%. Two additional \ot/H$\beta$ data points \citep{Barker1979, Bandy2023} that were not in \citet{Arkhipova2014} confirm this mild evolution from what is already a high \ot/H$\beta$ ratio of $\sim$10. 

Several other candidates with 50-60 year observational 
baselines were noted by \citet{Hajduk2015} including M\,1-11, M\,1-12, M\,1-26, H\,2-48 and H\,2-25 but these require more detailed investigation. 

There is more evidence for flux changes in cooler post-AGB stars (pre-PNe), where variations are seen in Hen 3-1357 \citep{Reindl2017},  CRL\,618 \citep{SanchezContreras2017} and IRAS~18062+2410 \citep{Cerrigone2017} but with some uncertainty whether this is secular evolution, event-driven evolution (e.g., post-thermal-pulse or common envelope) or variability.

\section{The planetary nebula IC\,418}  

IC\,418 (informally known as the Spirograph Nebula), is a well-known, high-surface brightness, compact, mildly elliptical PN 
in the constellation of Lepus. It measures 14\arcsec\ $\times$ 12\arcsec\  across \citep{Frew2016}. Two concentric but very 
faint ionized optical halos are better seen in the mid-infrared \citep{2012MNRAS.423.3753R}.   The ionized region has an 
estimated mass of 0.06\,M$_\odot$ \citep{Morisset2009}.  A photo-dominated region (PDR) with an estimated mass of $\sim 
0.5\,\rm M_\odot$ contains neutral hydrogen  \citep{Taylor1989} and other atomic lines \citep{Liu2001}. A dusty, molecular 
region shows carbonaceous dust, including PAHs and weak fullerene bands \citep{Otsuka2014,DiazLuis2018}. The nebula is carbon-
rich, with C/O\,$\approx 1.3$ \citep{Liu2001}. 


The distance to IC\,418 is well constrained.  \citet{Guzman2009} determined a radio data expansion parallax distance of $1.3\pm0.4$\,kpc, while \citet{Morisset2009} presented a photo-ionization-model distance of 1.26\,kpc.  The best statistical distance is $1.3\pm0.3$\,kpc from a well determined optical
surface-brightness radius relation \citep{Frew2016}. The Gaia DR3 parallax is $0.73\pm0.03$\,mas, 
yielding $1.36\pm 0.05$\,kpc which agrees well with these other estimates. We adopt the robust Gaia distance. The 
luminosity in Table~\ref{basic} in Appendix-B is scaled from the original reference to this Gaia distance.

Our reddening estimate, from an average of 12 independent determinations of the H$\alpha$/H$\beta$ ratio, 
is $E(B-V)=0.21\pm 0.06$. The 6\,cm radio flux over the Balmer line flux gives a reddening of $E(B-V)=0.19$ 
assuming $R_V=3.1$. A recent MUSE observation finds $E(B-V)=0.18$ based on the Pa9 and H$\beta$ line ratio 
\citep{Monreal2022}. The asymptotic reddening in this direction is $E(B-V)=0.20$ \citep{Schlafly2011}. All values are in good agreement.  IC\,418 is 600\,pc above the Galaxy's mid-plane, 
well above the clumpy interstellar dust layer, so agreement between the derived extinction and the asymptotic extinction is not unexpected. There is some evidence for additional internal extinction of $E(B-V)=0.05$ in the PN's bright rim \citep{Monreal2022}.

The integrated H$\alpha$ flux is $F({\rm H\alpha}) = -9.02\pm0.04$\,\ergcms\ from \citet{Frew2013}, based on an 
average of two separate measurements; the de-reddened value is $I({\rm H\alpha}) = -8.81\pm0.0$5\,\ergcms.   
\citet{Ventura2017} argue for a low metallicity with [O/H] of $-0.6$ but photo-ionization models of the nebula indicate solar-like values \citep{Morisset2009, Dopita2017}, with some evidence for depletion of Mg and Si (0.5~dex) and a large depletion of Fe (2.9~dex). We adopt the abundances of \citet{Morisset2009}.
 
\section{The central star}
The CSPN of IC\,418, HD\,35914, is bright and of spectral type O7f
\citep{Heap1977}, with a temperature reported as
$T_{\rm eff} = 33.5$\,kK \citep{Dopita2017}, 36.7\,kK
\citep{Morisset2009} or 39\,kK \citep{Escalante2012}, derived from
photo-ionization models fitted to modern observational data. 
\citet{Dopita2017} considers the CSPN to have evolved from a fairly
massive carbon star with initial mass of 2.5--3\,M$_\odot$.

Stellar magnitude measurements for HD\,35914 often include
contributions from the compact PN, unless explicitly corrected.
\citet{Ciardullo1999} give $V$ = 10.23 from HST data (after correcting
for nebular contribution) while Whole Earth Telescope data from
\citet{Handler1997} finds $V\simeq9.88$ which is corrected for nebular
light. Several recent photometric observations with APASS (the AAVSO
Photometric All-Sky Survey) and with UCAC4 \citep{UCAC42012}, give
values 0.5 to 1.2 mag brighter due to inclusion of nebular light given
the PN is of high surface brightness. Gaia DR3 photometry gives a
B-band magnitude (from 4000 to 5000\AA) of 9.879 and R-band magnitude
(from 6000 to 7500\AA) of 9.810.


The reported stellar luminosity of $L\sim7500\pm1000\,\rm L_\odot$ 
places it on the constant luminosity horizontal part of the post-AGB evolutionary tracks \citep{Bertolami2016}, 
with a core mass of $\sim 0.57$--$0.62\,\rm M_\odot$ and an initial mass of $\sim 1.5$--$2.5\,\rm M_\odot$ based on those 
theoretical tracks, which is somewhat lower than the \citet{Dopita2017} estimates.
  
The CSPN shows both small-amplitude photometric and wind variations in modern  observations and is the prototype of a class 
of cool, variable CSPN \citep{Handler1997, Kuczawska1997}. The variability is $\sim$0.2 mag level peak-to-peak about 
the current mean value, over time scales of hours. The $B-V$ colour similarly varies by 0.05 mag. The wind velocity is 
$v_w=500\,\rm km\,s^{-1}$ \citet{Morisset2009}. The star has an enhanced helium abundance, with He/H$\approx 0.25$ 
\citep{Morisset2009}. The nebula does not show this, indicating that stellar abundances have been modified since the ejection 
of the nebula.

The photographic DASCH \citep{DASCH2009} archive provides long-term photometric data for HD\,35914. This shows a $\sim$0.2 B-band dimming contributions from the evolving H$\beta$ and \ot\ PN nebular lines within the blue band-pass. Only the 497 unflagged (higher quality) DASCH points were used but these still have considerable scatter. A linear fit quantifies the dimming rate as 0.0026~mag/year. The UCAC4 B magnitude from epoch 1985.94 is 9.405 c.f. the DASCH trend line value at the same epoch of 9.505 in reasonable agreement. The nebular contribution leaves uncertainty in the amount of stellar dimming.

The basic observed and derived properties of PN IC\,418 and its CSPN are summarised in Table~\ref{basic} in Appendix~A.
\\

\section{Secular Evolution of the PN IC\,418}
The  nebula expansion has been measured directly using high-resolution radio observations \citep{Guzman2009} over 20 years. They derive a radial expansion rate of $5.8\pm1.5\, \rm mas\, yr^{-1}$.  \citet{Schoenberner2018} analyse HST images over  8 year timescales and find that the nebula expands by a factor of $8\times 10^{-4}$ per year, or $5.5\pm0.5\, \rm mas\, yr^{-1}$, in good agreement with the other measure.  \citet{Schoenberner2018} derive a PN age of $\sim1185\pm110$~years while \citet{Morisset2009} derive an expansion age of 1400~years. Regardless, IC\,418 is among the younger PNe known. Taking a current stellar $T_{\rm eff} = 37$\,kK, and an end-of-AGB temperature of $\sim 5$\,kK,  the average heating rate of the star over the entire time interval since the star left the AGB becomes $\sim$20--30\,K\,yr$^{-1}$. Such a rapid heating rate may cause measurable ionization structure evolution of the photo-ionized nebula over decadal timescales. This is confirmed by our investigations here.

\subsection{The historical spectroscopic archive}

Below we present a table of our evaluation of the historical records of optical spectroscopy of IC~418 that 
cover the \ot\ and H$\beta$ lines, supplemented by our own observations. Full details are given in Appendix~B.

\begin{table*}
\scriptsize
\addtolength{\tabcolsep}{-1pt}
\begin{center}
\caption{Secular change in the observed generally integrated \ot$\lambda$5007/H$\beta$ ratio with time.}\label{ratios}
\begin{tabular}{llccccl}
\hline
Date     &  \ot/H$\beta$  & Line Measures & Telescope & Instrument & Detector & Reference		\\
\hline						
 1893.8	&  $0.8\pm0.3$ 	& integrated	&  Lick 36" refractor & slitless spectrograph & eye & \citet{Campbell1893}  \\	
  1899	&  $<1.0$ 	& integrated	&  Lick 36" refractor & slitless spectrograph & eye & \citet{Keeler1899}  \\	
1916.8--1917.8 & 0.965* & integrated & Lick 36" refractor  & slitless spectrograph & photographic & \citet{Wright1918} $^1$\\
1938.7       &  1.39   	& integrated  & Lick 36" refractor & slitless spectrograph &  photographic & \citet{Aller1941}  	\\	
1940.9 & 1.35: & integrated &  Lick 36" refractor & slitless spectrograph & photographic & \citet{Wyse1942} $^2$ \\	
 1952.5 & 1.38 & integrated & 24" McMath-H Obs. & interference filters & photoelectric & \citet{Liller1955} \\
 1953.8 &  1.33 & integrated & Curtis Schmidt & slitless & photoelectric & \citet{Liller1954} $^3$ \\
 1959.5          &  1.38  & slit 30.4\arcsec wide & 36" Pine Bluff & scanning spectrograph & photoelectric & \citet{Capriotti1960} $^4$ 	\\	
 1959--1963 & 1.27 & integrated & 50cm Shternberg & slitless & photographic & \citet{Vorontsov1966SvA} \\
1962.8 & 1.27 & integrated  & 60"/100" Mt. Wilson & scanning spectrograph &  photoelectric & \citet{ODell1963} $^5$ 
\\
1966.9 & 1.37: & slit 10\arcsec $\times$45\arcsec & Lick 120" reflector & Coude spectrograph & imagetube+Ilford & \citet{Aller1970} \\
 1967.5$\pm$1.5   &  1.32 & integrated & Lick Crossley 36" & aperture & photoelectric  &  \citet{Peimbert1971}  	\\		
 1972--1978  & 1.47 & 2\arcsec$\times$2\arcsec slots & Lick/Mt Wilson & apert.spectra 2 posns & image tube & \citet{Aller1979} $^6$ \\
 1972.2$\pm$1.5 &  1.53  & slit 8\arcsec & Hamilton Crossley~36" & sequential scanner	& photoelectric & \citet{Barker1978}   	\\
 1973--1975 & 1.70 & slit 4.8\arcsec$\times$71.5\arcsec & CTIO &image tube spect. & IIDS & \citet{Torres1977}\\			
1974.5$\pm$1.5       		&  1.53  & integrated	& SAAO 0.5m & interference filters & photometer & \citet{Webster1983}  	\\	
 1976--1978.3 &  1.55   & integrated & ESO 0.5/1.0m & interference filters	& photoelectric &  \citet{Kohoutek1981}  	\\		
 1980.5       		&  1.51  & integrated & CTIO 0.9m & spectral scanner &	photomultiplier		&  \citet{Gutierrez1985}		\\
 1982.8-1983.8 & 1.52 & integrated & ESO 1.52m & B\&C spectrograph & IDS & \citet{Louise1984} $^7$ \\
 1984.3--1985.3   &  1.64  & integrated & CTIO 0.9m  & interference filter & aperture phot.			&  \citet{Shaw1989}	$^8$\\
 1986.8 & 1.52 & slit 4\arcsec$\times$4\arcsec & ESO 1.52m & B\&C spectrograph & CCD & Unpublished Acker $^{9}$\\
 1989--1991 & 1.54  &  slit 4\arcsec~ EW & Brazil 1.6m & B\&C spectrograph & Reticon & \citet{Freitas1992} \\
1992.2 &  0.86  &  slit 6\arcsec localised& Shane 3m & Echelle & CCD & \citet{Hyung1994} $^{10}$ \\
2006.0       		&  1.75  	& integrated & KPNO 0.6m & Fabry-Perot & CCD  & This work [WHAM] \\			
2008.2       		&  1.84  	& integrated& 3.9m AAT	& SPIRAL IFU & CCD	&  This work [AAT] \\
2008.4       		&  1.75  	& slit 2\arcsec~ EW	& 2.3m MSSSO &  DBS & CCD	&  This work  [MSSSO] \\
 2014.1 & 1.84 & integrated & ESO 8m VLT & MUSE IFU & CCD & Private communication $^{11}$\\
 2016.0             &  1.99    & integrated    & 2.3m MSSSO & WiFeS IFU & CCD             & \citet{Dopita2017}\\
 2016.9 & 2.09  & slit 2.5\arcsec~ EW & SAAO 1.9m & Cass Spec & CCD & This work [SAAO] \\
 2019.8 & 2.01 & integrated & 2.3m MSSSO & WiFeS IFU & CCD             & This work [MSSSO]\ \\
 2024.8 & 1.95 & slit 2.7\arcsec~ EW & SAAO 1,9m & SpupNic & CCD & This work [SAAO]\\ 
 \hline
\end{tabular} \\
\end{center}
* The formal lower limit from reported line intensities is 0.64, however, the high blue response of the emulsion is not corrected for. This  would increase this value to 1.29 but this is an upper limit. Taking the average gives 0.965\\
$^1$ The Wright spectrograms show H$\beta$ stronger than \ot\ but the \ot\ 5007/4959 ratio is reported as 1.57 c.f. the 
theoretically expected value of 2.98. This indicates a strong, blue sensitivity term for the photographic emulsion so the 
raw \ot/H$\beta$ line ratio of 0.63 is a clear lower limit.
$^2$ The intensity ratio in the paper is an eye estimate \citep{Aller1964} albeit done with with
the aid of specially prepared scale plates. Our line ratio is based on the 4959\AA\ line as the reported 
5007/4959\AA\ ratio is unphysically large while close to the theoretical value for the other 6 PNe reported 
in the paper. Hence, we have assumed that the \ot~5007\AA~ value of 230 is a typographical error and should be 130 
which puts it on near par with all the other measurements in the paper; 
$^3$ From the reported sum of the two \ot\ lines, assuming a line ratio of 3.0; 
$^4$ The paper gives no observing date but the observatory only opened in 1958, so the observations 
were likely done in 1959 or early 1960 so we selected 1959.5;
$^5$ The \ot/H$\beta$ line ratio used is scaled from the \ot\ 4959\AA~ line. Fluxes appear low for [Ne\,{\sc iii}] but ok for [O\,{\sc 
ii}]. The slit was likely positioned off-centre so the result is likely less representative. 
$^6$ Average of two measurements given. No date is given for the observations but the instrument was in use from 1972; $^7$ Average 
of the central 12 values in map for \ot\ and H$\beta$. No dates of observation were provided but the telescope was built in 1981; $^8$ The \ot/H$\beta$ line ratio is scaled from the \ot\ 4959\AA\ line.
$^9$ Unpublished spectrum from Stenholm and Acker uploaded to HASH; the \ot/H$\beta$ ratio is measured directly from 
their 1-D spectra where the line peaks give 1.52 and integrated value gives 1.51; $^{10}$ Spectrum is from a localised 
region towards the PN edge and is not representative of the PN as a whole, it is shown as a lower limit; $^{11}$ integrated 
ratio value determined from MUSE images of \ot\ 4949\AA~ and H$\beta$ then scaled to \ot\ 50007\AA\ equivalent from 
ESO 8m VLT MUSE IFU reduced data provided by Ana Monreal Ibero (private communication).
\end{table*}

\subsection{Spectroscopic evolution}
 \begin{figure*}
\begin{center}
\includegraphics[width=17cm]{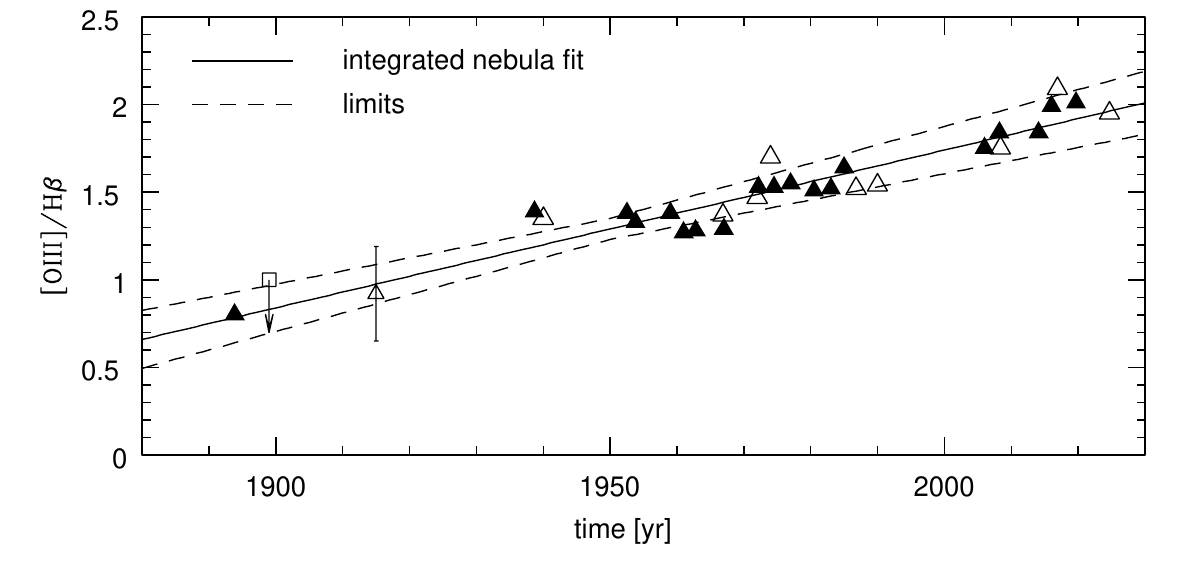}
\caption{Observed values of $\lambda$5007/H$\beta$ over a 130 year
  time period. The most reliable data, based on integrated flux across
  the full PN, are plotted as filled triangles. Open triangles refer
  to less reliable measurements usually derived from wide slits or
  apertures that do not cover all of the nebula. Open squares show
  limits. The off-centre data of \citet{Hyung1994} is not included.
  The best fit and limits are shown; this fit excludes the data from
  narrow slits (Table \ref{ratios}).}
\label{fig:flux_time}
\end{center}
\end{figure*}

The observed \ot/H$\beta$ ratio is plotted against time in Fig. \ref{fig:flux_time}.  The mostly integrated line ratios show 
a clear secular increase. IC\,418 is compact and mildly elliptical with a major axis of 14\arcsec\ so even long-slit spectra 
with typical 2-3\arcsec\ slits combined with typical 2-4\arcsec\  seeing at the modest observing sites used implies that a 
reasonable fraction ($\sim20$\%) of the full PN is sampled with slit spectra. A simple analysis of the AAT SPIRAL IFU data 
presented from 2008 shows that extracting virtual 2\arcsec\ wide slices across the PN through the centre in $15\deg$ radial 
increments only yields a variation of $\leq7$\%  in the \ot/H$\beta$ ratio and is in good agreement with the observed, integrated ratio. 

Accepting the \citet{Campbell1893} visual value, the \ot\ $\lambda$5007 line has more than doubled in strength compared to
H$\beta$ since the discovery of the PN. This clearly indicates a change in photoionization. \citet{Dopita2017} identify a 
shock contribution to this line of around 2.5\% but this is much smaller than the observed increase. 

We fitted the observed line ratios with a linear fit, using the python routine {\tt linregress} from {\tt scipy} version 
1.13.1. Selecting only the entries in Table \ref{ratios} covering the entire nebula (integrated or slit wide enough to cover 
the full nebula: filled symbols in Fig. \ref{fig:flux_time}) gives: 
\begin{eqnarray}
\nonumber F(\lambda5007/{\rm H}\beta) &=& (0.0090\pm0.0015) \,\times (t-1950) \\
 & &+ (1.29\pm0.06) ,
\label{eq:1}
\end{eqnarray}
 
\noindent where $t$ is the calendar year. The uncertainties were derived by adding normal noise to the data points, 
assuming $\sigma=0.3$ for the 1893 Campbell datum and $\sigma=0.1$ for all other data points. The 1899 Keeler and 1917 
Wright data are not used in this fit (but are still plotted). 

Including all data points in Table \ref{ratios} apart from \citet{Aller1979} and \citet{Hyung1994} gives:
\begin{eqnarray}
\nonumber F(\lambda5007/{\rm H}\beta) &=& (0.0089\pm0.0012)  \,\times (t-1950) \\
& &+ (1.30\pm0.05) .
\label{eq:2}
\end{eqnarray}
 

Without the Campbell datum, the slope becomes $(0.0091\pm0.001$\,yr$^{-1}$ in Eq.\,(\ref{eq:1}) and $0.0089\pm0.0008 
$\,yr$^{-1}$ in Eq.\,(\ref{eq:2}). Thus, the Campbell value has little effect on the slope itself but dominates the 
uncertainty. 

We conclude that the \ot/H$\beta$ ratio has increased approximately linearly by 0.9\%\ per year. Continued monitoring will show whether this trend continues.

\begin{figure*}
\begin{center}
\includegraphics[width=13cm]{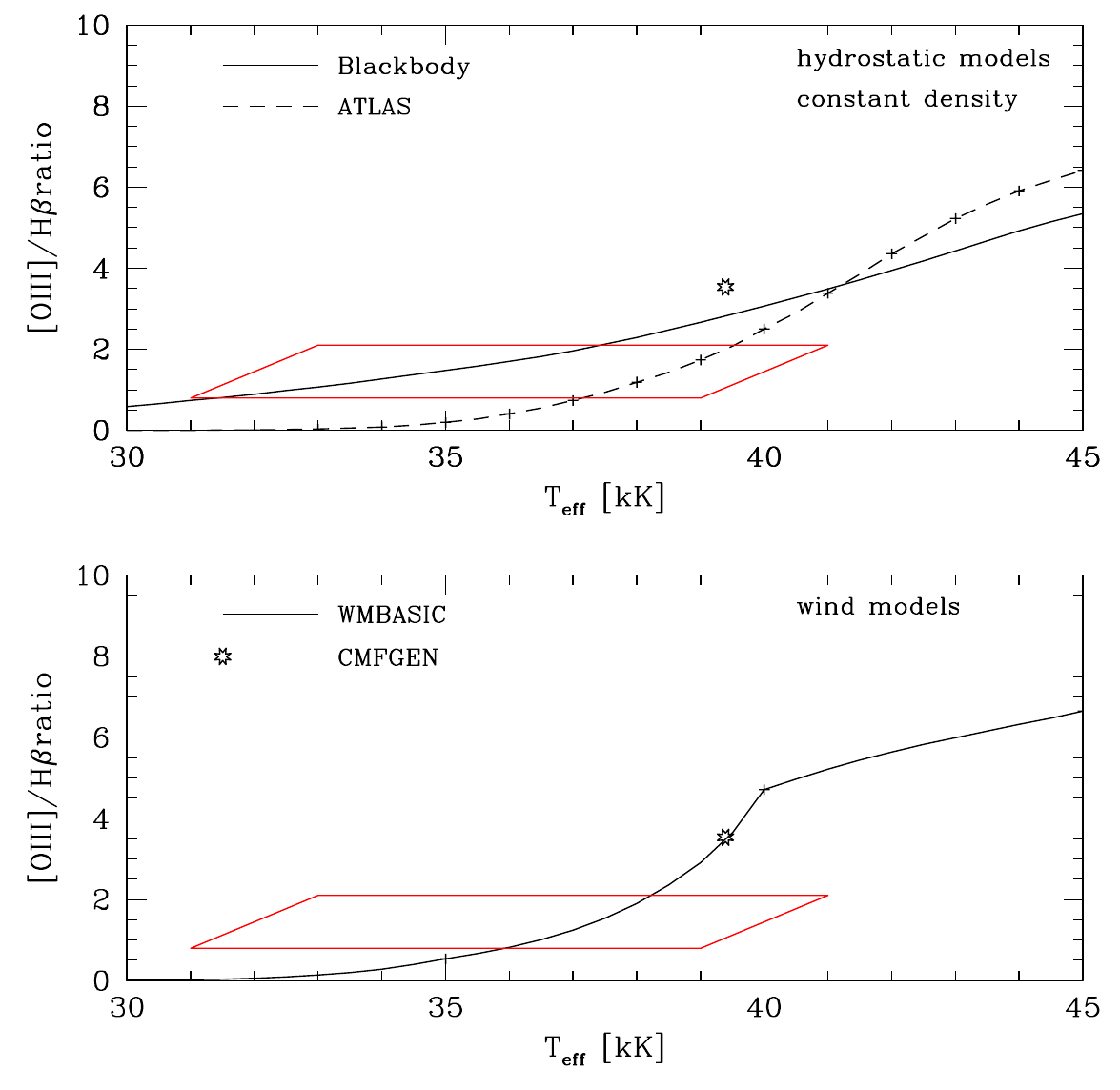}
\caption{The predicted \ot\ 5007/H$\beta$ ratio as function of $T_{\rm eff}$, for different stellar atmosphere 
models and assuming a constant density nebula. Top panel: Hydrostatic ATLAS and blackbody models. Bottom panel: WMBASIC   models which include stellar winds, and  the Cloudy fit for a single CMFGEN model with stellar winds from \citet{Escalante2012} (also shown in the top panel).  The red parallelogram encompasses the range of observed \ot\ 5007/H$\beta$ flux ratios. The plusses indicate the stellar models which Cloudy uses for interpolation. 
\label{fig:fluxratio_temp}}
\end{center}
\end{figure*}

\section{Modeling}

The observed line ratio change is caused by the star heating
up. Modeling the nebula and star is required to determine their
relation. We aim not to reproduce the detailed spectrum but to model
the relative changes in the \ot/H$\beta$ ratio. The mild (1-arcsec)
angular expansion of the nebula over the historical evolution is also
not considered.

A comprehensive model fit to the IC\,418 PN spectrum was given by
\citet{Morisset2009} that included an extensive imaging and
spectroscopic study via a pseudo~3-D photoionization code. We use a
simplified 1-D version of their nebular model to investigate the
dependency of the integrated \ot/H$\beta$ flux ratio on stellar
temperature. The \citet{Morisset2009} model combines 1-D models
calculated for four different directions across the PN. This provided
a good approximation while avoiding the computational complexity of
full 3-D modeling. We reduced this to a single 1-D model, adopting the
parameters of \citet{Morisset2009} for the equatorial plane.  For a
CSPN, a blackbody function gives correct line ratios for elements with
ionization potentials up to $\sim 40$\,eV \citep{Morisset2009}. The
O$^{++}$ ion requires 35.1\,eV to form, and the \ot\ lines have a
significant dependency on the stellar atmosphere model used.

For our modeling we used nebular abundances from \citet{Morisset2009}
and tested the abundances of \citet{Dopita2017} but with no
significant differences seen. Other parameters from
\citet{Morisset2009} were also adopted, but we use the Gaia DR3
distance of $1.360\pm0.055$~kpc: the inner radius, scaled to that
distance, is $\log r_{\rm in} = 16.09\,\rm cm$. A 1-D Cloudy model was
computed out to the ionization radius where the electron temperature
drops to $T_e=10^3$\,K.  Cloudy version 23.01 was used
\citep{Ferland2013, Chatzikos2023} but in 1-D mode rather than the 3-D
approximation used by \citet{Morisset2009}, equivalent to assuming a
spherical nebula. This is a reasonable approximation for this mildly
elliptical PN \citep{Escalante2012} with an axial ratio of 0.85. Shock
excitation is not included in Cloudy, but \citet{Dopita2017} find that
this contributes around 2\%\ to the \ot\
emission. \citet{Guerrero2013} also find little evidence of shock
emission in IC\,418.

The main models used a constant density, taken as
$\log n_e = 3.95 \,\rm cm^{-3}$. Both \citet{Morisset2009} and
\citet{Dopita2017} adopt a double-shell model, albeit with different
inner radii. We also computed models for both these distributions and
found some minor differences in the integrated \ot/H$\beta$
ratios. The \ot\ emissivity is dominated by the inner region, limiting
the effect of the radial density distribution.

Cloudy model grids were calculated for three stellar atmosphere models
provided in Cloudy: blackbody, ATLAS ODFNEW \citep{Castelli2003} at
$\log g=4.5$ and WMBASIC \citep{Pauldrach2001} at $\log g=4.5$. 
  TMAP models developed for CSPN \citep{Rauch2018} do not cover the
  temperature range for IC\,418 and could not be used, while CoStar
  models \citep{Schaerer1997} are for much higher stellar masses.
The ATLAS models are for hydrostatic, plane-parallel atmospheres, with
a temperature grid at steps of 1000\,K. The WMBASIC models are
developed for CSPN and include stellar winds. They are available at
temperatures of 35, 40 and 45 kK.  The value for $\log g$ for ATLAS
and WMBASIC was chosen as the closest available to the $\log g=3.55$
for IC\,418 within the model set which covered the required range of
stellar temperatures. All models used solar metallicity.  The single
CMFGEN model of \citet{Escalante2012} for their best fit temperature
was also kindly provided to us by C. Morisset (priv. comm.). This
model was also used in \citet{Gomez-llanos2018}.

The stellar luminosity is set at $\log L = 37.30\,\rm erg\, s^{-1}$ (5200 L$_\odot$) for the blackbody models and at $\log L = 37.40\,\rm erg\, s^{-1}$ (6500 L$_\odot$) for all other models. The lower $L$ for blackbody models is because this predicts a lower temperature and therefore brighter star at visual wavelengths for the same \ot/H$\beta$ ratio. 

The stellar temperature  was increased in steps of 500\,K, between limits of 30\,kK and 45\,kK, and the ratio  
of \ot/H$\beta$ over the entire nebula was obtained from the Cloudy models for each stellar temperature. 
The H$\beta$ flux increases by about a factor of 2 over this temperature range, mainly over the lower temperature range 
\citep{Zijlstra1989}, but the \ot\ flux increases faster giving a monotonic increase in the ratio over this temperature range.

The results from the four models with constant $n_e$ are shown in Fig.~\ref{fig:fluxratio_temp}, separated between the  
hydrostatic and the wind models. The \ot/H$\beta$ ratio increases strongly over the plotted temperature range, 
but the details vary between the models. For the same ratio, 
the derived $T_{\rm eff}$ can differ by $\sim$10\% especially at low temperatures. Blackbody models yield  higher 
\ot/H$\beta$ ratios at low stellar temperatures than do the other models, but at higher temperatures the others overtake them 
because of the effect of the ionization absorption edges, especially He$^+$ which has an ionization potential very close to 
that of O$^{2+}$. The CMFGEN model of \citet{Escalante2012} is represented by the open star symbol: it agrees with the WMBASIC model grid, but predicts an integrated \ot/H$\beta$ ratio well above the observed value and appears to overestimate the stellar temperature.

\begin{figure}
\begin{center}
\includegraphics[width=8cm]{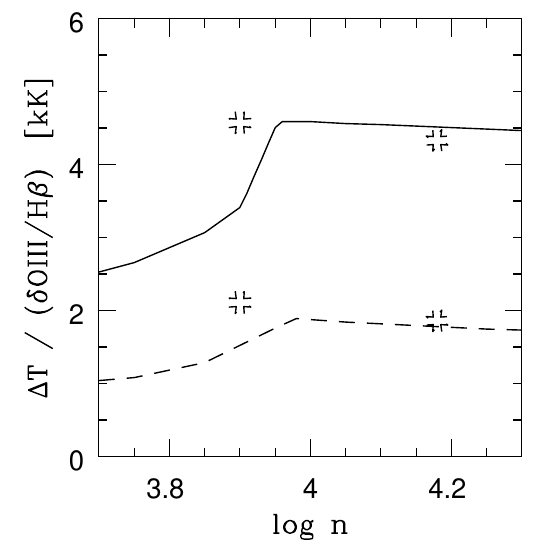}
\caption{The change in stellar temperature needed for increasing
  \ot/H$\beta$ from 1.0 to 2.0, for blackbody and ATLAS models at
  constant density.  The drawn line is for blackbody models and the
  dashed line for ATLAS stellar atmospheres.  The points indicate
    the location of the \citet{Morisset2009} (left) and
    \citet{Dopita2017} (right) density models, shown at the density of
    their main inner shell, where the two upper points are for
    blackbody SEDs and the two lower ones for the ATLAS SEDs. }
\label{fig:density}
\end{center}
\end{figure}

The non-constant density distributions of \citet{Morisset2009} and \citet{Dopita2017} yield minor changes. 
The two models differ in inner radius but this has no effect. They also have different densities. For densities below the 
selected value of $\log n=3.95$  and \ot/H$\beta <2.5$, the ratio increases faster with stellar temperature. For higher densities, 
the evolution is much less density dependent. Fig. \ref{fig:density} shows the dependency on density, where the vertical axis gives the change in temperature required to change the \ot/H$\beta$ ratios from unity to 2, based on Cloudy 
models. The lines are for constant density models; the points shows the location of the density models of \citet{Morisset2009} and \citet{Dopita2017}, plotted at the density 
of their inner shell.  For the former model, increasing the density by 0.05 dex does not notably change the value. We conclude 
that for the parameters of IC\,418  the spectral evolution is little affected by the precise density, but if the density had 
been significantly lower, then the \ot/H$\beta$ would have increased faster with stellar temperature. Note that the actual stellar temperatures do depend on density, even where the rate of change does not.

\begin{figure}
\begin{center}
\includegraphics[width=8cm]{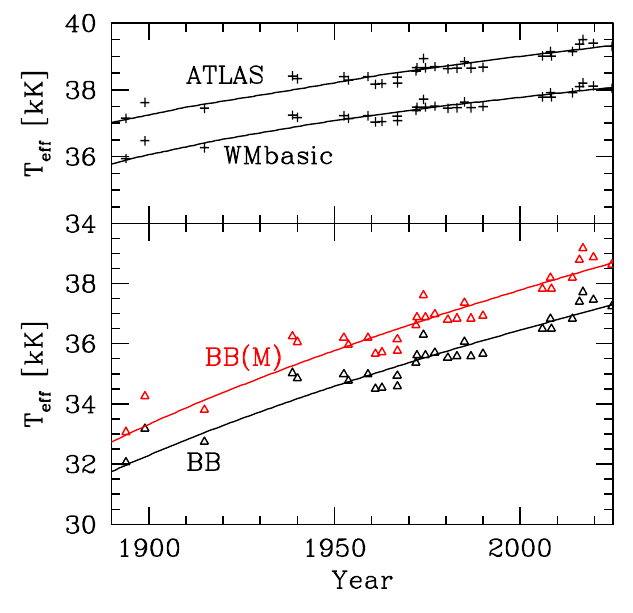}
\caption{The derived central star temperature evolution of IC\,418 based on changes in the observed \ot/H$\beta$ line ratios. The drawn lines correspond to the linear fit of Eq. \ref{eq:2}. The lower panel  gives blackbody SEDs from the density distributions of \citet{Morisset2009} in red - labelled BB(M) - and \citet{Dopita2017} - labelled BB - respectively.}
\label{fig:temperature}
\end{center}
\end{figure}

\section{Stellar evolution}
\subsection{Heating rate }

To convert the observed line ratio to stellar temperature, we linearly interpolated the ratios to the Cloudy temperature 
grids.  This was done separately for the data points and for the linear fit of Eq. \ref{eq:2}. The results are shown in Fig. 
\ref{fig:temperature}. The scatter reflects the uncertainty on the measured line ratio and on the temperature. 
Around $T_{\rm eff}=37\,$kK,  a 5\%\ error on the line ratio translates to a temperature uncertainty of 100\,K for the model atmosphere but 300\,K for a blackbody. The 1893 measurement has a larger uncertainty in the ratio, and the associated temperature is less well determined with an uncertainty of 500\,K for the model
atmosphere and 1000\,K for the blackbody. 

We fit the data  to infer the  temperature rate of change assuming linear evolution. Including the  \citet{Campbell1893} visual data point gives:  
\begin{eqnarray}
\nonumber T_{\rm eff} &=& (34.54\pm 0.10)\times 10^3 + (38.8\pm2.5)\times (t-1950)\,\rm
                 K\\
  & &\quad \hbox{blackbody} \\
\nonumber T_{\rm eff} &=& (38.20\pm 0.04)\times 10^3 + (16.1\pm1.0)\times (t-1950)\,\rm 
                K\\
  & &\quad \hbox{ATLAS} 
\end{eqnarray}

\noindent where $t$ is the calendar year. The highest value for the slope is obtained for the \citet{Morisset2009} density distribution with a blackbody stellar atmosphere (shown in red and labeled 'BB(M)' in Fig. \ref{fig:temperature}):
\begin{eqnarray}
 \nonumber T_{\rm eff} &=& (35.72\pm 0.11)\times 10^3 + (41.7\pm2.7)\times (t-1950)\,\rm  K\\
\end{eqnarray}

The WMBASIC model is similar to the ATLAS result.
Excluding the \citet{Campbell1893} point reduces the slope by $\sim5\%$, which is within the uncertainty of the fit. The slope is
little affected by inclusion of the earliest data point. 

The  uncertainty on the conversion from \ot/H$\beta$ evolution to a heating rate is dominated by the stellar atmosphere, which gives 
a range of a factor of four. The density structure causes additional uncertainty but for the parameters of IC\,418 this is 
less than 10\%. Based on the explored parameter range, the highest heating rate is 41.7\,K/yr, based on the density 
distribution of \citet{Morisset2009} and a blackbody atmosphere. The ATLAS stellar atmospheres give a heating rates of 
$16.1\pm 1.0$\,K/yr and WMBASIC $15.2\pm1.0$\,K/yr.  

\subsection{Stellar mass}
The heating rate is a strong function of central star mass \citep{Gesicki2014}. Recent post-AGB models of 
\citet{Bertolami2016} ($Z=0.01$, H-burning) predict heating rates ranging from 2 to 180\,K/yr at stellar temperatures of 30--40\,kK, for central star masses between 0.53 and 0.62\,M$_\odot$. These heating rates  cover two orders of magnitude. The  luminosities vary only by a factor of $\sim 3$ over this mass range. This makes the heating rate by far the best determinant of stellar mass.

The derived model-dependent heating rates of 10--40\,K/yr corresponds to stellar masses of 0.560--0.583\,M$_\odot$, where the stellar atmosphere 
models are at the low end and the blackbody models at the high end of the mass range. Both lower and higher masses can be excluded at good confidence. The indicated mass range is also consistent with the stellar luminosity derived here, of $L=5200$-$6500$\,L$_\odot$.

The stellar mass is related to the initial stellar mass by the initial-final mass relation. Using the initial masses in the models of \citet{Bertolami2016}, the range corresponds to main-sequence masses of $M_{\rm i} =1.25$--$1.55\rm \,M_\odot$.

An interesting aside is that the initial-final mass relations show a bump around initial masses of 1.5-1.9\,M$_\odot$ \citep{Marigo2020} which is apparent in the \citet{Bertolami2016} models. The heating rates do not show this and appear to correlate better with the initial masses than do the final masses.

\section{Discussion and conclusions} 

The heating rate provide a narrow mass range for IC\,418's CSPN. The nebula is carbon rich \citep{Liu2001}. This provides 
another constraint as low-mass stars do not dredge-up sufficient carbon to achieve C/O$>1$. \citet{Marigo2020} find a minimum 
mass of $M_{\rm i}>1.65\,\rm M_\odot$ for stars to become carbon stars at solar metallicity, with C/O$\,=1.3$ (the value for 
IC\,418 from \citet{Liu2001}) reached for $M_{\rm i}=1.8$--$1.9\,\rm M_\odot$.  These values are outside the mass range 
derived from the heating rate. In contrast, \citet{Rees2024} find a minimum initial mass for carbon star formation of $M_{\rm i}=1.5$--$1.75 \,\rm M_\odot$, and \citet{Karakas2016} derived 1.4--2\,$M_\odot$ which does overlap with the mass range derived here but only for blackbody models. The stellar atmosphere models, which may be more realistic, give masses that are below the carbon-star range. This important contradiction indicates that some aspects of the AGB-PN evolution may be uncertain.

Given the complexity of this evolutionary phase, it is difficult to pin-point one aspect that might be the issue. A possibility is in the stellar atmosphere models which are not optimised for post-AGB stars; it is important to compute new models which include a wind, as is present in IC\,418 \citep{Morisset2009}. Using other ions than O$^{++}$ may reduce the sensitivity to these models. 

Alternatively, the post-AGB stellar evolution models may overestimate heating rates.  Instantaneous heating rates derived 
here are compatible with those derived from the  nebula's kinematic age ($\sim 25$\,K/yr), leaving the possibility that post-
AGB models evolve too fast. They also use grey atmospheres for the calculations rather than full atmosphere models, so the 
model temperatures may deviate from the observed ones. Finally, the initial-final mass relations may overstate the final masses, as possibly indicated for NGC\,3132 \citep{DeMarco2022} and cluster PNe \citep{Fragkou2025}. 

In conclusion, we have, for the first time, shown robust, direct, secular evolution of a CSPN over an unprecedented 130 year 
time period. This provides an important new tracer for the evolution of post-AGB stars. Implications are that at solar metallicity the lower mass cutoff for carbon-star formation may need further consideration.


 
\newpage
\section*{Acknowledgements}

Dr. David Frew initiated this project and carried out research across Sections 1–4, but unfortunately has left academia. He deserves special mention: the authors gratefully acknowledge his crucial role.

Q.A.P. thanks the Hong Kong Research Grants Council for GRF research support under grants 17304024, 17326116 and 17300417. A.A.Z. thanks the Hung Hing Ying Foundation for a visiting HKU professorship and acknowledges UK STFC funding under grants 
ST/T000414/1 and  ST/X001229/1. A.A.Z. also acknowledges support from the Royal Society through grant IES/R3/233287 and 
from the University of Macquarie. This work made use of the University of Hong Kong/Australian Astronomical 
Observatory/Strasbourg Observatory H-alpha Planetary Nebula (HASH PN) database, hosted by the Laboratory for Space Research 
at the University of Hong Kong. Use was made of the TMAW tool (http://astro.uni-tuebingen.de/~TMAW), constructed as part of the 
activities of the German Astrophysical Virtual Observatory. The authors thank Dr. Andreas Ritter for assistance with AAT 
SPIRAL, ANU 2.3m WIFES IFU and SAAO 2024 SpuPNIC data of IC\,418. 
  We thank Ana Monreal Ibero for the ESO 8m VLT MUSE IFU data 
point from 2014, and Christophe Morisset for providing us with the CMFGEN model for the central star.

\bibliographystyle{aasjournalv7}

\bibliography{IC418}

\appendix

\section{Basic Properties of PN IC\,418 and its central star HD~35914 from the literature or this work}
\label{sec:appendix_data}

\begin{table}[h]
\scriptsize
\begin{center}
\label{basic}
\begin{tabular}{lrl}
\hline
Name                           	& IC\,418   			&	HASH ID \#752 \citet{Parker2022}		\\
\hline
RA (J2000)                   	& 05 27 28.21  			& 	\\				
Dec (J2000)	          	& $-$12 41 50.3		& \\

$l$ 	                       		&    215.212   	 		&			\\				
$b$ 	                       		& $-24.284$  			&			\\	
Distance                            & $1.36\pm0.055$~kpc		& Gaia DR3 		 \\
\hline
{\it Planetary Nebula}			& 					&			\\			
Dimensions (\arcsec)  	& 14 $\times$ 12 		&	\citet{Frew2016}			\\
log F(H$\alpha$)               & $-9.02\pm0.04$   		& \citet{Frew2013}  \\				
log F(H$\beta$)                 & $-9.58\pm0.05$ 		& \citet{Acker1992}   \\
$F_{\rm 6cm}$                   & 1710 mJy  			& \citet{Griffith1994}\\			
$c$                                    & $0.31 \pm 0.08$     	& this work 		  \\
$E(B-V)$                           & $0.21 \pm 0.06$     	& this work 	 		 \\
$\left<n_{\rm e}\right>$                                & 9000\,cm$^{-3}$ 		& \citet{Dopita2017} \\
$T_{\rm e}$                                & 9000\,K  			& \citet{Morisset2009} \\
Inner radius                       & $1.23\times 10^{16}$\,cm   & 	\citet{Morisset2009}	\\
Outer radius                      & $1.4 \times 10^{17}$\,cm    & 	\citet{Morisset2009}	\\
                                  &   0.045~pc  & 		\\
Kinematic age                   & $\sim1185\pm110$ yr          & \citet{Schoenberner2018}\\
                  & $\sim$1400 yrs          & \citet{Morisset2009} \\
\it{v}$_{\rm exp}$              & 30km/s & \citet{Guzman2009} \\
\hline
{\it Central star* } 				& 					&  				\\
$V$                                	& 10.23  		 		& \citet[][HST]{Ciardullo1999}	\\
$I_c$                               	& 10.22  		 		& \citet{Ciardullo1999}		\\
$T_{\rm eff}$  			&  36.7\,kK  			& \citet{Morisset2009} \\
$L$                                	& 7500\,$L_{\odot}$  	& \vtop{\hbox{\strut \citet{Morisset2009},}\hbox{\strut 
\citet{Krticka2020},} \hbox{\strut \citet{GM2024}}} 
\\  
$\log g$ &  3.55 & \citet{Morisset2009} \\ 
Spectral type               &  O7f & \citep{Heap1977}\\
Wind velocity  $v_w$       &   $500\,\rm km\,s^{-1}$    & \citep{Morisset2009}\\
$M_{\rm core}$              &  0.560--0.583\,$M_{\odot}$                & this work         \\
$M_{\rm progenitor}$         &  1.25--1.55\,M$_{\odot}$           & this work         \\
\hline
\end{tabular}
\end{center}
*nebular contribution removed from photometric estimates of the star c.f. tabulated values in SIMBAD where V=9.01 from 
UCAC4 data of the system \citet{UCAC42012}.
\end{table}

\section{The historical spectroscopic archive}
\label{sec:appendix_archive}
To assess evidence for secular evolution in line intensity ratios and of the diagnostic \ot/H$\beta$ ratio, 
it is essential to carefully evaluate the integrity of all available spectral observations reported for IC\,418 over the 
130 year time period of available data. This is important given the wide range of technologies employed where 
variables include: observatory sites and telescopes (refracting and reflecting), spectrographs 
(slits, slitless, prisms, grisms, gratings, IFUs and Fabry-Perot etalons) and detectors (eye, photographic plates, 
image tubes and CCDs). We have carefully examined all original publications and, where justified, applied 
corrections based on subsequent knowledge (such as the now known theoretical line ratio for the \ot\ doublet). 
This was objectively done to provide the best possible values for this study. The results are summarised in Table~\ref{ratios} 
in the main body of the paper, with particular attention paid to the observations prior to the advent of CCDs.

IC\,418 was discovered in 1891 by Williamina Fleming at Harvard Observatory \citep{Pickering1891}. 
The \ot/H$\beta$ ratio was noted as being notably low compared to other PNe known at the time. She stated that 
"the visual spectrum differs strikingly from that of other planetary nebulae". In 1891 IC\,418 would have been 
$\sim$0.7\arcsec\ (5\%) smaller than it is now based on the modern determined expansion rates \citep{Guzman2009, Schoenberner2018}.

\citet{Campbell1893, Campbell1894} appears to have made the first recorded spectral observation on Nov~2, 1893, 
with a visual spectroscope with a wide slit to include all the PN. He noticed that the two ``nebulium'' lines 
N1 (4949\AA) and N2 (5007\AA) (unidentified then, but now known to be the \ot\ lines) were slightly less extended 
and more centrally condensed than H$\beta$.  He measured a brightness ratio by eye using a dark wedge and blocking off an 
increasing fraction of the light for one line until the brightness appeared equal. This showed the brightness of the H$\beta$ 
line to lie midway between the two \ot\ doublet lines. He quotes a brightness ratio for $\lambda$5007:H$\beta$:$\lambda$4959 
of 10:7:3. Multiplying these surface brightness values by the respective image sizes (which he reports as 11 and 14\arcsec) 
gives an intensity ratio of $\lambda$5007:H$\beta$ of $0.9\pm0.3$.  We correct this for the scotopic spectral response 
of the human eye using the CIE 1951 curve \citet{CIE1951}  
which at H$\beta$ is 88\%\ that at \ot~5007\AA.  The corrected ratio is \ot/H$\beta = 0.8\pm0.3$.  

The fact that the [\ot\ image is more 'compact' than H$\beta$ was confirmed by \citet{Keeler1899} but 
Keeler does not provide intensity ratios. Based on the perceived size difference we report his \ot/H$\beta$ value  as $<1$. 

Subsequent observations were made with photographic plates, image tube scanners or photoeletric photometers, 
sometimes in combination, until the advent of CCDs from 1976 onwards, e.g. \citep{1976ccdt.conf..135S, 
2015PASP..127.1097L}. The earliest photographic observation appears to be from \citet{Wright1918} who obtained blue spectra 
from a quartz slitless spectrograph that appeared to also show H$\beta$ stronger than \ot, again implying that 
\ot$\lambda$5007:H$\beta<1$. However, there is clearly some strong non-linearity in the blue response of these early Eastman Kodak "Seed 23" glass plate emulsions as a function of wavelength. 
This is apparent from the reported relative intensities of the two ``nebulium'' lines (i.e. \ot5007/4959\AA) 
are given as 33:22, a ratio of only 1.5 when this ratio should theoretically be 2.98. The formal reported 
$\lambda$5007:H$\beta$ relative intensity ratio is 33:51 (ratio 0.64) which is a clear lower  
limit. Simply correcting the H$\beta$ line for the observed to the theoretical \ot5007/4959\AA\ ratio gives a  
$\lambda$5007:H$\beta$ ratio of 1.29. This is likely an over-correction so we regard this as the upper limit. 
We adopt a mid-point value of $0.965\pm0.3$ as the best estimate from these data.

No spectra were published for the next 20 years until results from a careful analysis of a quartz 
slitless spectrograph photographic plate spectra from the Lick Observatory Crossley 36-inch reflector in 1938 \citet{Aller1941}. Here the \ot\ line was reported brighter than H$\beta$ with a formal reported ratio of 1.39. There is some confidence in this value as the reported relative intensities from the \ot\ 5007/4959\AA\ for IC\,418, and indeed for 6 other PNe in his paper, have values close to the theoretical value of 3.0 with an average for all the PNe of 2.83 with $\sigma = 0.09$.
Subsequent measurements found \ot/H$\beta$ $\approx1.4$. All post year 2000 measurements continue to give increasing ratios that are now reported in excess of 1.7 and up to $\sim2.0$. 

The most reliable \ot/H$\beta$ ratios collected and assessed are given in Table \ref{ratios}.  
The exact observation time is not always known, and some papers give a range of dates. This is reflected in 
the uncertainties on years given in the table.  Some measurements did not integrate along the slit but present 
ratios from the PN's centre where \ot\ is stronger due to ionisation stratification \citep[e.g.][]{Torres1977}.  
The deep spectrum of \citet{Hyung1994} was centred on the outer ring where \ot\ is weak. These  measurements are 
listed but are excluded from the analysis. Some papers use earlier observations to supplement or calibrate their data and so cannot be used as independent measurements 
\citep[e.g.][]{Osterbrock1961,Aller1970,Boeshaar1974}. Hence these estimates were excluded from our analysis.

Measurements derived from photographic plates are more uncertain due to calibration difficulties, especially as 
the \ot\ 4959 and 5007\AA\ lines fall at the far red end of the sensitivity curve for the old blue emulsions. 
This affects the \citet{Aller1941} results, who observed IC\,418  with the Crossley reflector and slitless quartz 
spectrograph at Lick Observatory in September 1938, among a group of about a dozen PN. Five IC\,418 plates were 
taken with the deep blue Wratten 50 filter and Eastman 33 emulsion of the so called “photographic region” with 
exposures of 5 to 31 minutes. The Wratten filter has very low transmission redward of 5000\AA~ (Wratten filters 
4th edition, Eastman Kodak Co. 1920) but both \ot\ 4959 and 5007\AA~ line intensities were recorded for all PNe in 
the sample. The 5007/4950 ratios from the reported line intensities range from 2.69 (NGC\,6572) to 2.96 (NGC\,6826), 
c.f. the theoretical ratio of 3.0. The value for IC\,418 is 2.78 from the 
reported intensities of 13.9 and 5.0 – the weakest for all PNe in the sample.

\begin{figure*}
\begin{center}
\includegraphics[width=18cm]{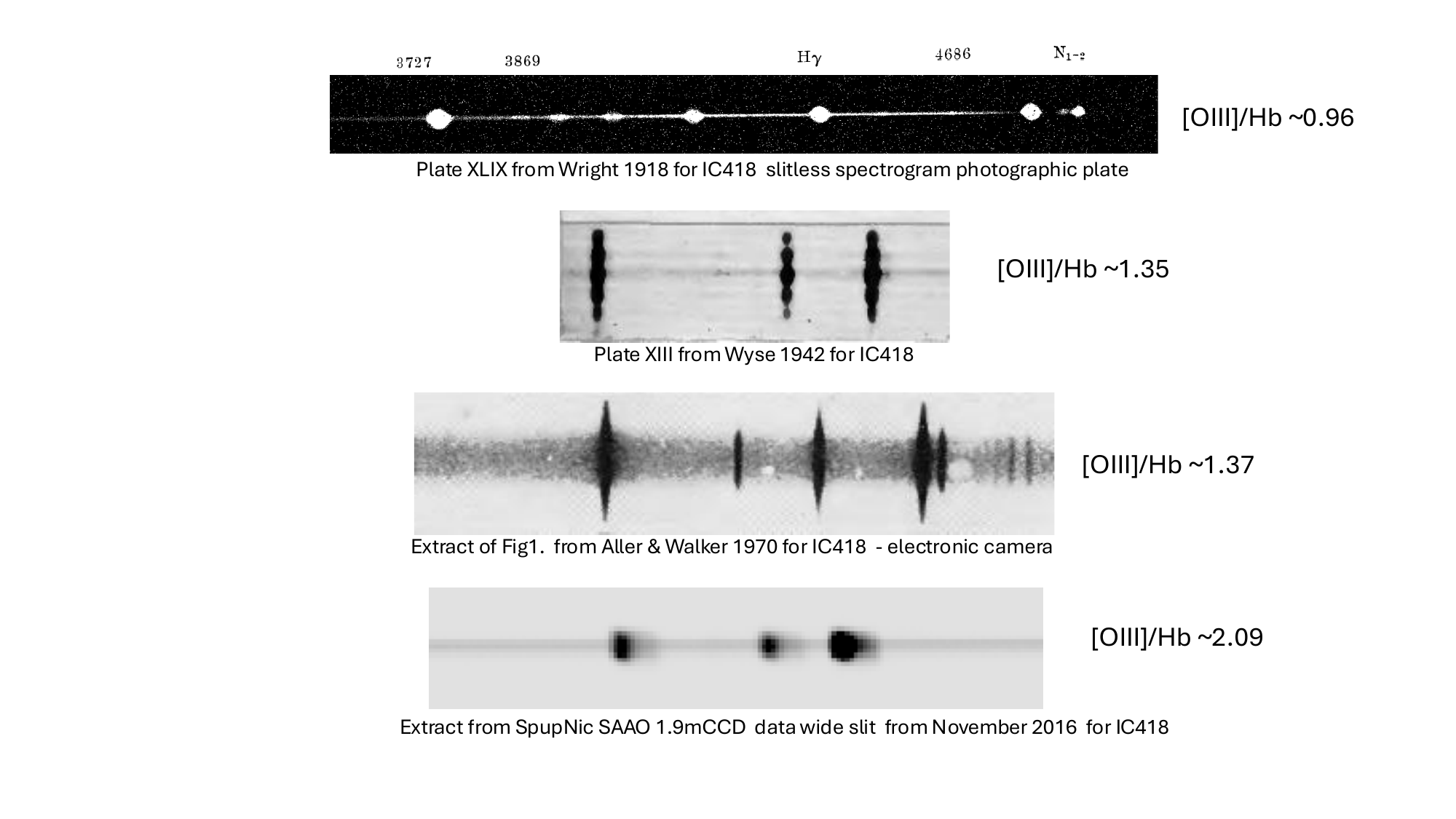}
\caption{Selected historical examples of 2-D spectra of IC418 covering the H$\beta$ and \ot\  region from years 1918 to 2016. The change in the \ot\ to H$\beta$ ratio is clear, notwithstanding corrections for emulsion sensitivity for the 1918 and 1942 photographic plates. From top to bottom we show a slitless IC\,418 spectrum given in \citet{Wright1918}; a slit spectrum from \citet{Wyse1942}; an electronic camera spectrum from \citet{Aller1970} and finally a 2-D extract from the CCD SpupNic image from the SAAO 1.9m taken by the second author in November 2016 all together with the estimated \ot\ to H$\beta$ ratios.}
\end{center}
\end{figure*}

\label{lastpage}
\end{document}